# Ultra-fast GHz-range swept-tuned spectrum analyzer with 20 ns temporal resolution based on a spin-torque nano-oscillator with a uniformly magnetized "free" layer

*Artem Litvinenko [1,\*], Ahmed Sidi El Valli [1], Vadym Iurchuk [1], Steven Louis [2], Vasyl Tyberkevych [2], Bernard Dieny [1], Andrei Slavin [2] and Ursula Ebels [1]*

[1] Univ. Grenoble Alpes, CEA, CNRS, Spintec, 38000 Grenoble, France

[2] Oakland University, 48309 Rochester MI, USA

**The advantage of an ultra-fast frequency-tunability of spin-torque nano-oscillators (STNOs) that have large (> 100MHz) relaxation frequency of amplitude fluctuations is exploited to realize ultra-fast wide-band time-resolved spectral analysis at nanosecond time scale with the frequency resolution limited only by the "bandwidth" theorem. The demonstration is performed with an STNO generating in the 9 GHz frequency range, and comprised of a perpendicular polarizer and a perpendicularly and uniformly magnetized "free" layer. It is shown that such a uniform-state STNO-based spectrum analyzer can efficiently perform spectral analysis of frequency-agile signals with rapidly varying frequency components.**

**Introduction**

Performing spectrum analysis with sub-µs time resolution is expected to bring significant progress in the field of signal processing and data communication [1-3]. This can be achieved with a swept-tuned spectrum analyzer, only if the local oscillator frequency can be swept on the time scale of the targeted time resolution. In a recent work [4] a novel approach to achieve fast spectrum analysis was demonstrated experimentally. In that work, the key component of a high-speed spectrum analyzer was a spin-torque nano-oscillator (STNO) [5]. In comparison to traditional oscillators, the STNOs bring several advantages to fast spectrum analysis, such as a nanoscale size, rapid frequency tuning by an external DC bias current, and a possibility of seamless integration with semiconductor integrated circuits [4, 6-14]. For spectrum analysis, it is important that nano-sized STNOs can be tuned very fast. Their tuning speed is limited, mainly, by the rather high relaxation frequency of their amplitude fluctuations [6, 7, 10-16].

In a preliminary study [4], fast spectrum analysis was performed using a magnetic tunnel junction (MTJ) STNO with a "free" layer in a vortex ground state operating at a frequency below 500 MHz, and having an amplitude relaxation frequency $f_p$ of the order of 1 MHz [4, 15, 16]. In that initial study, spectral analysis was demonstrated with the time resolution of 1µs and the capability to follow the time evolution of an input signal frequency. However, for a real breakthrough in practical applications the time resolution of spectral analysis needs to be reduced to a time resolution well below the 1µs limit.

To do this, we propose to use an STNO based on a uniformly magnetized nanopillar MTJ that generates microwave signals at much higher frequencies, and characterized by much higher relaxation frequency of amplitude fluctuations [10, 11]. This is non-trivial, as often the uniformly magnetized MTJs, compared to the vortex-state MTJs, are characterized by a higher intrinsic phase noise and lower output power. For comparison, a vortex-state STNO may have an output power of the order of 1µW, and a generation linewidth of about 250 kHz [15-17]. In contrast, the best corresponding values for STNOs based on uniformly magnetized MTJs are about 0.2 µW and 4 MHz [18, 19]. This last result was achieved with a crossed configuration of an in-plane polarizer and a perpendicularly magnetized "free" layer. It is important to note, that both the vortex-state and the uniformly magnetized STNOs demonstrated a reduction of the generation linewidth when a phase-locked loop was used in the oscillator design [19, 20].

In order to demonstrate fast spectrum analysis at the sub-µs time scale, we have developed a novel type of STNO, which has a linear frequency-voltage dependence in the range of 8.6-9.6 GHz, a generation linewidth of 35 MHz, and a relaxation frequency of amplitude fluctuations $f_p$ above 300 MHz. This STNO is comprised of a perpendicular polarizer and a perpendicularly magnetized "free" layer. In the following we, first, evaluate the generation performance of this novel STNO type, and then, demonstrate the use of this STNO for ultra-fast spectrum analysis at sweeping rates $\rho_{sw} = \Delta F_{sw}/T$ exceeding 60 MHz/ns in a frequency interval of $\Delta F_{sw} = 1$ GHz, allowing us to achieve efficient microwave spectrum analysis with temporal resolution on the ns time scale.

**Results**

**STNO devices**

The STNO devices developed for this experimental demonstration of spectrum analysis are magnetic tunnel junction nanopillars of the structure "Pol/MgO/FL", where Pol stands for polarizing layer, FL stands for free layer, and the spacer layer is magnesium oxide (MgO).

The polarizer design is derived from perpendicular magnetic memory structures, using a thin $Fe_{72}Co_8B_{20}$ layer that is sandwiched between the MgO tunnel barrier and a (Co/Pt) multilayer. With this design, the Pol has a uniformly out-of-plane magnetization, see Fig. Fig. 1(a). Its structure is, from bottom to top, as follows: Pt25/(Co0.5/Pt0.25)$_{11}$/Co0.5/Ta0.3/FeCoB1.5, with the number after the material reporting the thickness of the layer in nanometers, and the subscript representing the number of repetitions of the Co/Pt multilayer. The FL has a structure that is an inverted version of the polarizer, with a FeCoB layer thickness that is slightly thicker than the one of the polarizer, so that it would have an in-plane magnetization without exchange coupling to the biasing (Co/Pt) multilayer. With this the FL structure is: FeCoB2/Ta0.2/(Co0.7/Pt1.1)$_{10}$/Ta4. The MgO thickness and oxidation was adjusted to a nominal value of the resistance area (RA) product RA = 10 $\Omega\mu m^2$. The materials were sputter-deposited and nanofabricated using e-beam lithography, ion beam etching, as well as chemical etching techniques to form circular elements with a diameter of 100 nm, or elliptical elements with dimensions of 100x120 nm. A major hysteresis loop obtained with an out-of-plane field, typical of the fabricated devices, is shown in Fig. 1(b). When in zero field after negative saturation, the FL-FeCoB layer and the Pol-FeCoB layer are both aligned with a negative field, shown as "downward pointing red and blue arrows" in the figure. As the field increases, there is a gradual increase of the resistance between 0 Oe and about 3 kOe that is interpreted as a gradual tilt of the FL-FeCoB layer away from the negative saturation. Between 3 and 4 kOe, the resistance jumps into the anti-parallel (AP) state where the FL-FeCoB layer reverses to align with the external field, while the Pol-FeCoB layer remains anti-aligned with the external field. For these STNOs, the best linewidth and linear frequency tuning were observed when the device was in the AP state with an out-of-plane field of ~ 4 kOe, and the in-plane device shape was elliptical with dimensions of 100 x 120 nm.

For the RF measurements, a permanent magnet was used to produce an out-of-plane field of about 4 kOe with a small in-plane tilt angle between 1° and 5°. Typical results of the free running signal are shown in Fig. 1(c) for an elliptical device characterized by a parallel (antiparallel) resistance value of $R_p \approx 110$ Ω ($R_{AP}$=118 Ω), a tunneling magnetoresistance value of TMR≈7% , and a measured RA≈5 Ω µm².

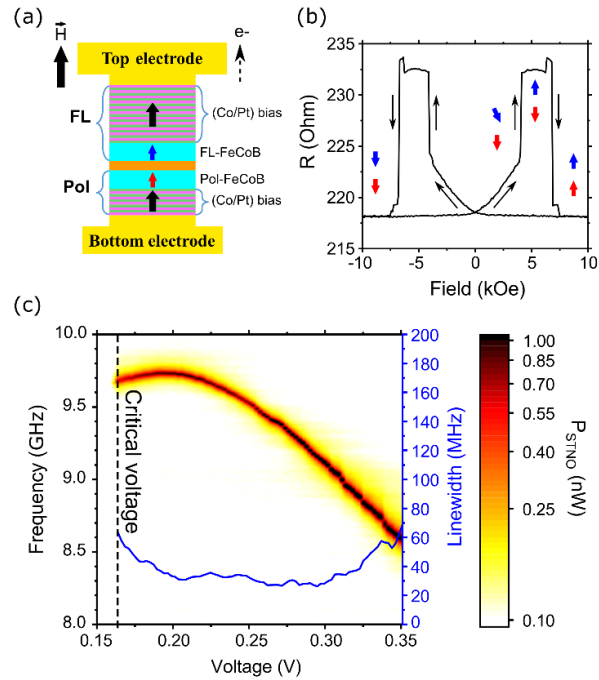

**Figure 1** (one column figure): (a) Schematics of the STNO comprising a perpendicularly magnetized polarizer multilayer and a perpendicularly magnetized "free" multilayer, as described in the text. In the absence of the bias multilayer, the "free" layer would be magnetized in-plane. The role of the bias multilayer is to pull the "free" layer magnetization out-of-plane via exchange coupling; the black dashed arrow represents an electron flow; the black arrow shows the external field direction which is out-of-plane. (b) Typical magnetoresistance curve of the fully perpendicular uniformly magnetized STNO under the out-of-plane bias field; (c) Typical frequency-voltage characteristics $f_o(V)$ measured with an out-of-plane bias magnetic field of $H_\perp$= 4 kOe. Well defined steady state excitations of the free layer are observed for positive applied voltage, corresponding in our sign convention to electrons flowing from the polarizer to the "free" layer. Note, that all measurements presented in Figs. 2-4 were carried out for the same device as the one shown in (c). Experiments on other devices gave similar results.

For a bias field of 4 kOe and an applied voltage that sweeps from 0.2 to 0.35V (1.7-3 mA), the frequency generated by the STNO varies from 9.6 to 8.6 GHz. As can be seen, these STNO devices reveal a frequency red shift with a remarkably high slope and linear tuning of 10GHz/V (or 0.77GHz/mA) in the range of 0.25-0.35V. The corresponding full width at half maximum (FWHM) linewidth is $\Delta f$ = 35 MHz (corresponding to a Q-factor $f/\Delta f$ of 260), and a power about 1.5nW. The low power is due to the low TMR value of 7%. It is expected that with device optimization to TMR= 100%, this device would yield the output power of at least 0.1μW. The relaxation frequency of amplitude fluctuations in the developed STNO devices is evaluated from modulation experiments to be larger than 300 MHz. The relatively low linewidth of this uniformly magnetized STNO configuration as well as the large frequency tuning range of $\Delta F_{sw}$ =1 GHz, are what make the demonstration of ultra-fast wideband spectrum analysis possible.

**Spectrum Analyzer scheme**

The experimental scheme employed for the STNO-based spectrum analysis is shown in Fig. 2(a). Details about this experimental scheme can be found in [4]. Below, we describe the essential steps necessary to generate a swept-tuned reference signal $V_{ref}(t)$ and point out the differences from our previous study. First, a constant bias DC voltage $V_{dc}$ = 0.3 V is applied to the STNO to generate a sinusoidal auto-oscillation signal with a frequency of $f_0$ = 9.1 GHz. This DC voltage was chosen so that $f_0$ lies in the center of the linear frequency tuning range, as shown in Fig. 1(c).

The bias voltage is modulated, via a coupler, by a triangle wave form $V_{sw}$ that continuously sweeps the STNO frequency up with a sweep time of T and down with a sweep time of T, as shown in Fig. 2(a)(i). Using the triangular shaped signal has an advantage over the sawtooth wave form used in the previous study, because this wave form eliminates the abrupt jumps, and, thus, can provide higher sweep frequencies, while avoiding generation of higher harmonics. Also, both the "up" and "down" sweeps can be used for the signal spectral analysis, so that the corresponding sweep frequency is $f_{sw} = 1/T$. The amplitude of $V_{sw}$ is adjusted so that the frequency of the auto-oscillation signal $V_{STNO}$ generated by the STNO is swept between the two frequency values $f_1 = 8.6$ GHz and $f_2 = 9.6$ GHz around the central frequency of $f_0 = 9.1$ GHz with a span of $\Delta F_{sw} = f_2 - f_1 = 1.0$ GHz. The signal $V_{STNO}$ (t) is, then, passed through an amplifier and a high-pass filter with a cut-off frequency of 1.0 GHz to produce a reference signal $V_{ref}$. The reference signal $V_{ref}$ has a frequency chirp with the instantaneous frequency $f_{ref}(t)$ that varies with time, as illustrated in Fig. 2(a)(iii).

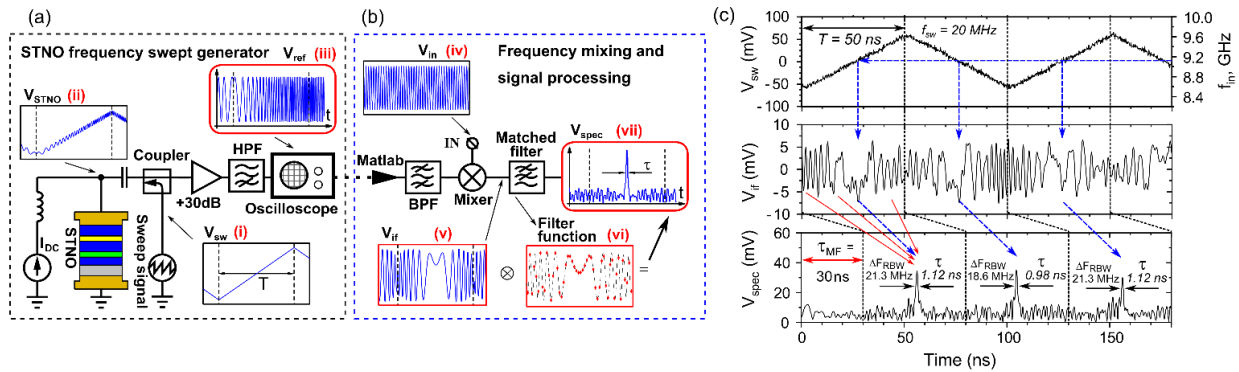

**Figure 2** (two column figure): Schematics of the ultra-fast wide-band spectrum analyzer based on an STNO with a uniformly magnetized "free" layer. Block (a) illustrates the generation of a swept-tuned reference signal $V_{ref}$, while block (b) illustrates the signal procession implementation in "MATLAB", resulting in the spectral analysis of the incoming test signal $V_{in}(t)$. The insets show the (noiseless) representative voltage signals vs. time at different points of the spectrum analyzer circuit: (i) the triangular sweep signal $V_{sw}$, (ii) the output of the STNO $V_{STNO}$, (iii) the reference signal $V_{ref}$ produced by the swept-tuned STNO after amplification and filtering, (iv) the external test signal $V_{in}$, (v) the mixed output signal $V_{if}$, (vi) the discretized basis filter function of the matched filter signal, (vii) the output signal $V_{spec}$ showing a peak of duration $\tau$, the temporal position "$t_0$" which identifies the information about the frequency of the analyzed external signal $V_{in}$, and the width of which determines the resolution bandwidth of the frequency analysis. (c) Spectral analysis of a single-tone external sinusoidal signal ($f_{in}$ = 9.1 GHz, $A_{in}$ = 1 V): (top panel) sweeping voltage signal $V_{sw}$, (middle panel) voltage $V_{if}$ and (bottom panel) voltage $V_{spec}$ for $f_{sw}$ = 20 MHz measured at the different points of the spectrum analyzer setup (see insets (i), (v) and (vii) in Fig. 2(a)). The experimental frequency resolution $\Delta F_{RBW}$ was calculated using $f_{ref}(t)$, and the experimentally measured duration $\tau$ of the peak seen in the voltage $V_{spec}$ (see bottom panel).

The generated time traces are, then, registered using a 50 GS/s oscilloscope. Subsequent signal processing (mixing and spectrum analysis of an external signal $V_{in}$ using a matched filter) has been performed within a MATLAB environment using the Simulink package as detailed below. In our previous work using low frequency carriers (300 MHz), signal processing was performed with a matched filter implemented in an FPGA. Unfortunately, an FPGA operating in the 10 GHz range was not available for this experiment, and thus, the signal processing was performed numerically, which does not compromise the validity of the concept and/or the obtained results.

The PC-based signal processing procedure is performed as follows. First, the reference signal $V_{ref}$ is digitized by the oscilloscope. Then, the signal $V_{ref}$ is passed through a band pass filter [8.6-9.6] GHz within the MATLAB program in order to improve the signal to noise ratio (SNR). Then, a numerically generated external signal $V_{in}$ of an unknown frequency $f_{in}$ is introduced. After that, the signals $V_{in}$ and $V_{ref}$ are mixed in a numerically modelled mixer. The modelled mixer had the output cut-off frequency of 2 GHz, and, therefore, performs the function of a low-pass filter with a bandwidth of 2 GHz, and, because of this, retains the mixing output component having the frequency $f_{in} - f_{ref}$, while removing the output component with the frequency $f_{in} + f_{ref}$.

The mixed signal $V_{if}$ having the frequency $|f_{in} - f_{ref}|$ is, then, passed through a matched filter having 830 taps (which is implemented as an FPGA-compatible finite impulse response (FIR) model) to compress this signal into a narrow peak with a duration $\tau$ in the voltage $V_{spec}$. The temporal position $t = t_0 + \tau_{MF}$ of the maximum of the peak in the voltage $V_{spec}$ allows the determination of the frequency $f_{in}$ of the input signal. Note, that this is similar to the case of ref. [4] when a real (analogue) matched filter was used, and requires to take into account an additional time delay of $\tau_{MF} = 30$ ns when mapping the temporal position $t_0$ into the frequency axis using the function $f_{ref}(t)$. The temporal width of this peak $\tau$, when mapped on the same axis, characterizes the frequency resolution bandwidth (RBW), or the accuracy of the frequency analysis: $\Delta F_{RBW} = \rho\tau = \Delta F_{sw}\frac{\tau}{T}$.

An example for the analysis of a monochromatic sinusoidal single-tone signal $V_{in}$ ($f_{in} = 9.1$ GHz, amplitude 1 V) is shown in Fig. 2(b). The panel at the top of the figure shows the sweeping triangle voltage $V_{sw}$ having a frequency $f_{sw} = 1/(T) = 20$ MHz. The middle panel shows the chirped signal mixed with $V_{in}$ resulting in the intermediate frequency signal $V_{if}$, and the bottom panel shows the output signal from the matched filter $V_{spec}$. From the bottom panel of Fig. 2(b), the temporal position of the resulting peak $t_0 = t - \tau_{MF} \approx 27$ ns (corresponding to $V_{sw} = 0$) yields the external signal frequency of $f_{in} = 9.1$ GHz.

We note, that while both $V_{sw}$ and $V_{if}$ are noisy, the action of the matched filter dramatically improves the signal-to-noise ratio, since it performs correlation processing of the signal. Basically, the signal hidden in the noise is squeezed into a narrow, but high-amplitude peak, while the uncorrelated noise passes through the filter unchanged. The typical peak in the lower panel of Fig. 2(b) has a characteristic duration of about $\tau = 1.12$ ns, which corresponds to a frequency resolution bandwidth $\Delta F_{RBW} \sim 21$ MHz at the used scan rate.

**Resolution bandwidth**

To determine how the frequency resolution bandwidth of an STNO-based spectrum analyzer depends on the sweep rate, the experiment was repeated for different sweep frequencies $f_{sw} = 4 - 50$ MHz. Typical time traces for different sweep rates are shown in Fig. 3(a), along with the average peak duration τ and average frequency resolution bandwidth $\Delta F_{RBW}$.

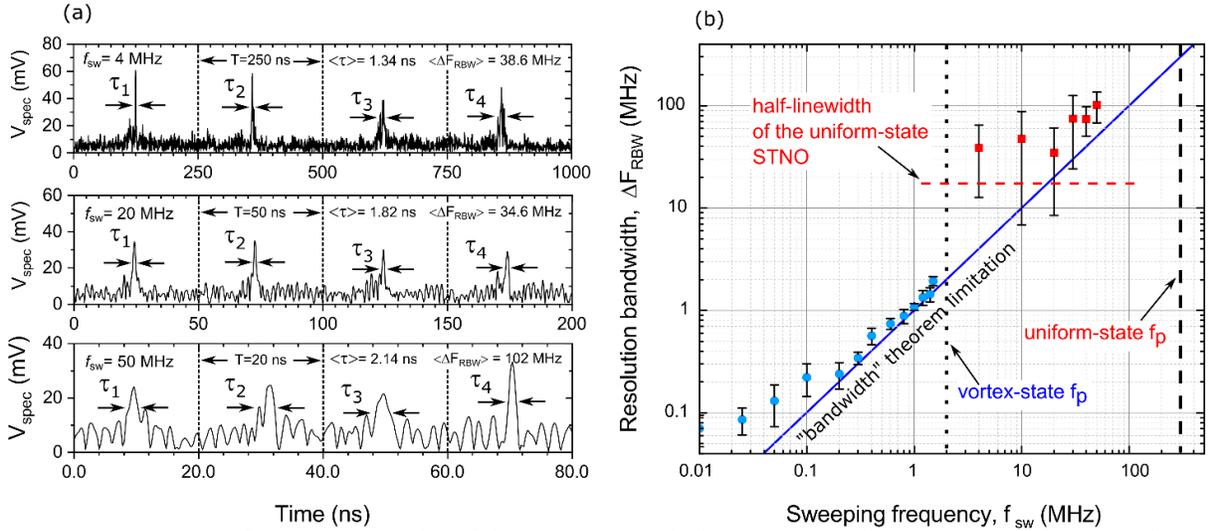

**Figure 3** (two column figure): (a) Results of the spectrum analysis (Vspec(t)) of a monochromatic external signal performed using a uniform-state STNO with the same parameters as in Fig. 2 (b) for three different sweeping frequencies fsw = 4, 20, and 50 MHz; (b) Resolution bandwidth (RBW) $\Delta F_{RBW}$ of the spectrum analysis as a function of the sweeping frequency fsw obtained using the average peak duration τ and the "sweeping" function fref(t). Red squares correspond to the uniform-state STNO, while light blue dots correspond to the vortex-state STNO [4]. Error analysis was done on 10 random consequent measurements of the peak duration τ. The straight dark blue line in Fig. 3(b) was calculated from the "bandwidth" theorem: $\Delta F_{RBW} = f_{sw} = 1/T$. The dashed and dotted vertical lines show the values of the experimentally determined relaxation frequency $f_p$ of amplitude fluctuations for the uniform-state and vortex-state STNO correspondingly (see [4,7,10] for details). The horizontal dashed line shows the magnitude of the generation half-linewidth of the uniform-state STNO.

It is important to emphasize, that the bottom panel of Fig. 3(a) shows the spectrum analysis performed on a 1 GHz bandwidth that is scanned in just 20 ns. In comparing this study of a uniformly magnetized STNO with a vortex-state STNO as in the previous study [4], here spectrum analysis is performed 50 times faster than before (20 ns vs. 1 μs), while sweeping a bandwidth that is 50 times larger (1 GHz vs. 20 MHz).

The resolution bandwidths obtained in this analysis for different sweeping rates are shown in Fig. 3(b) with red squares. A blue line also shown in this plot indicates the theoretical limit obtained from the "bandwidth" theorem $\Delta F_{RBW} = f_{sw} = 1/T$, while the blue points show the results from the previous study performed using a vortex-state STNO [4].

The results presented in Fig. 3(b) constitute the principal results of this paper. They show that a uniformly magnetized STNO can perform spectral analysis at microwave (> 5 GHz) frequencies with a frequency resolution that approaches the theoretical limit imposed by the bandwidth theorem. It is pertinent to determine the upper limit of the sweep frequency. In particular, it is important to understand when the STNO-based spectrum analysis fails to perform with the resolution defined by the bandwidth theorem. In other words, it is important to determine the minimum and maximum sweep times for a particular STNO.

From the perspective of the frequency resolution when using a matched filter, the minimum sweep time is constrained by the STNO generation linewidth. This is indicated in the figure by a red dashed line, which is at ½ the STNO generation linewidth. Thus, for lower sweep frequencies, the noise provides a floor for STNO based spectrum analysis frequency resolution. Likewise, the maximum sweeping frequency is determined by the intrinsic amplitude relaxation frequency $f_p$ that is much higher than the generation linewidth for the uniformly magnetized STNO device, as indicated by the vertical dashed line in Fig. 3b.

Within these two limits, the experimentally obtained frequency resolution bandwidth $\Delta F_{RBW}$ is very close to the theoretical limit defined by the "bandwidth" theorem (dark blue line in Fig. 3b).

This is an important result and means that even the relatively large generation linewidth of the uniformly magnetized STNO has practically no influence on the resolution bandwidth $\Delta F_{RBW}$ at high sweeping frequencies in the proposed method of ultra-fast spectrum analysis.

Here the maximum sweeping frequency that was achieved is much below the amplitude relaxation frequency and is determined by the parameters in the experimental procedure; specifically, by the combination of the analysed bandwidth and the sweep frequency.

For example, at a sweep frequency of 50MHz considering the theoretical resolution of 50MHz, for 1GHz of analysed bandwidth it would give only N = $\Delta F_{sw}/f_{sw}$ = 20 points of resolution for digital implementation of a matched filter. In order to improve the number of points the experiment would require either an increase of the analysis bandwidth or a reduced sweeping frequency. As N < 20 is not viable for real world applications, and 1 GHz is the maximum linear tuning bandwidth of this device, we did not evaluate the fastest sweep frequency above $f_{sw}$ = 50MHz.

**Signal with time varying frequency components**

Finally, we demonstrate that a spectrum analyzer based on a uniform-state STNO can successfully analyze complex external signals $V_{in}$, whose frequency content $f_{in}(t)$ is rapidly varying in time.

Results of this experiment are shown in Fig. 4, where $V_{in}$ has a frequency that varies in a "sawtooth" fashion (see Fig. 4(a)), i.e. the frequency increases linearly in time between $f_{min}$ = 8.9 GHz and $f_{max}$ = 9.3 GHz with a period of $T_{in} = 1/f_{sw}^{in}$ = 10 T (1/T = $f_{sw}$ = $10 f_{sw}^{in}$ = 20 MHz). The frequency of the external signal is shown in blue in Fig. 4(a). The black curve in Fig. 4(a) shows the triangular wave used to drive the STNO frequency sweeps. It is evident, that the STNO sweeps the entire scanning bandwidth multiple times while the frequency of the external signal is varying. The peaks in $V_{spec}$ obtained as a result of the spectrum analysis are shown in Fig. 4(b). The peaks in the signal $V_{spec}$ have different temporal positions, corresponding to different instantaneous frequencies of the external signal measured during consequent sweeping periods. The resultant temporal evolution of the frequency of $f_{spec}(t)$ is shown in Fig. 4(c). Thus, it is clear that a swept-tuned spectrum analyzer based on a uniformly magnetized STNO is also capable of not only unambiguous detection of the unknown frequency of the external input signal, but also of tracking of this frequency at a reasonably fast rate and with the frequency resolution close to the theoretical limit determined by the bandwidth theorem.

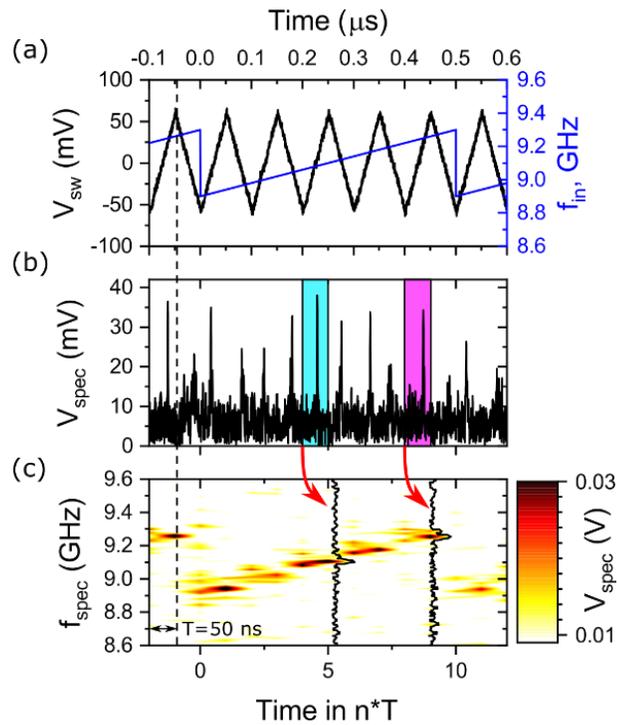

**Figure 4** (one column figure): Spectrum analysis, using the uniform-state STNO, of a complex external input signal $V_{in}$, whose frequency is varying in a sawtooth manner between 8.9 and 9.3 GHz with a period of $T_{in} = 10\ T$, where $T$ is the sweeping period $T = 50$ ns : (a) STNO sweeping voltage $V_{sw}$ (black line) and the time variation of the input signal frequency (blue line); (b) Resultant peaks $V_{spec}(t)$ with temporal positions varying within each sweeping period; (c) Temporal variation of the external signal frequency obtained as a result of the time-resolved spectral analysis. In the frame (c) the frequency axis corresponds to the frequency swept within a single period T of Fig.4 (b), and the color scale gives the peak height within a single sweep. Due to the triangle shaped sweep signal $V_{sw}$, subsequent sweeps alternate between frequency sweeping up and sweeping down. This was accounted for when compiling the results into the spectrogram shown in (c).

**Conclusions**

In this work we demonstrated that ultra-fast wide-band spectrum analysis with temporal resolution can be performed using a spintronic spectrum analyzer based on a uniform-state swept-tuned STNO driven by a triangular-shape bias current. The key elements of the STNO design is the perpendicular uniformly magnetized "free" layer in combination with a perpendicular polarizer. The developed STNO-based spectrum analyzer is able to analyze microwave signals in the frequency range of 1.0 GHz around a central frequency of $f_0$=9.1 GHz, with a scanning speed of 50 MHz/ns and the frequency resolution close to the theoretical limit defined by the bandwidth theorem.

The results demonstrate that the uniform-state STNOs, which often have much higher phase noise figures and larger generation linewidths, as compared to vortex-state STNOs, can be successfully used for the ultra-fast time-resolved spectrum analysis of complex frequency-agile dynamically modulated signals (see Fig. 4). At high sweep rates the frequency resolution $\Delta F_{RBW}$ of such ultra-fast spectrum analyzers becomes independent of the STNO generation linewidth $\Delta f$, and at the values of the sweep frequencies exceeding the STNO generation linewidth $\Delta f$ ($f_{sw} > \Delta f$) is limited mainly by the "bandwidth" theorem (see Fig.3 (b)).

We, however, would like to note that the use of the STNOs with narrower generation linewidth [5,21,22] would still be advantageous to improve the frequency resolution of spectrum analysis and to bring it even closer to the limit imposed by the "bandwidth" theorem for slow scanning frequencies.

AUTHOR INFORMATION

**Corresponding Author**

*Email.com: LitvinenkoAN@gmail.com

**Author Contributions**

A.L. and U.E. conceived the experiments; A.L. designed the experiment; A.L., A.S.E.V. and V.I. performed the measurements and analyzed the data; S.L., V.T., and A.S. performed theoretical calculations; A.L. prepared the samples; A.L., U.E., and A.S. wrote the manuscript with help from S.L. and V.I.; A.S., B.D. and U.E. managed the project; all authors contributed to the manuscript, the discussion, and analysis of the results.

**Notes**

The authors have no competing financial interests.

ACKNOWLEDGMENT

This work was supported in part by the ERC Grant MAGICAL (N°669204), by the Fondation Nanosciences, Grenoble, France, by the U.S. National Science Foundation (Grant # EFMA-1641989), by the Air Force Office of Scientific Research under the MURI grant # FA9550-19-1-0307, by the DARPA TWEED grant #DARPA-PA-19-04-05-FP-001, and by the Oakland University Foundation.

ABBREVIATIONS

STNO, Spin-torque nano-oscillator; MTJ, magnetic tunnel junction; RA, resistance area; AP, anti-parallel; FWHM, full width at half maximum; FPGA, Field Programmable Gate Array; SNR, signal to noise ration; FIR, finite impulse response; RBW, Resolution bandwidth.